\documentclass[12pt]{article}
\usepackage{graphicx}
\usepackage{amsmath}
\usepackage{amssymb}
\title{\bf Dirac's scalar field as dark energy within the frameworks of conformal theory of gravitation in Weyl--Cartan space}
\author{\bf Olga V. Baburova\footnote{E-mail: baburova@orc.ru},
Boris N. Frolov\footnote{E-mail: frolovbn@orc.ru}, Roman S. Kostkin\footnote{E-mail: kostkin@orc.ru},\\
{\em Moscow State Pedagogical University,}\vspace{-2.5mm}\\
{\em Department of physics for natural faculties,}\vspace{-2.5mm}\\
{\em MSPU, Krasnoprudnaya Street 14,}\vspace{-2.5mm}\\
{\em 107140 Moscow, Russian Federation}} \vskip 1cm
\date{}
\begin{document}
\maketitle
\begin{abstract}
Equations of the conformal theory of gravity with a Dirac scalar field in a Weyl--Cartan space-time have been derived. An exact solution of the equation for a scalar field, which has kind of a decreasing exponential function, has been found. This allows to explain a significant decrease (during of the period of inflation) the energy of physical vacuum (dark energy), which has been identified with the energy of the Dirac scalar field.  
\end{abstract}

According to calculations of the quantum field theory, values of the cosmological constant responsible for accelerated expansion of the Universe should differ on 120 orders at early and modern stages of evolution. In \cite{BKF:Kazan}, \cite{BKF:MathSb} one of the possible approaches has been developed to an explanation of such cosmological constant dynamics.

As the basis of the further constructions, we use Poincare--Weyl gauge theory of gravitation that has been developed in \cite{BFZ1} -- \cite{BFZ3}. This theory is ivariant both concerning the Poincare's subgroup and the Weyl subgroup  --  extensions and compressions (dilatations) of a spaces-time. In the theory there is the additional scalar field $\beta(x)$  introduced by Dirac. Its transformation under action of  dilatations looks like: $\delta\beta = \varepsilon (x)\beta$.

As a consequence, an invariance of the theory under the following conformal transformations arises:
\begin{equation*}
\begin{split}
&\delta g^M_{ab}=0\, ,\quad \delta g_{\mu\nu} = -2\varepsilon g_{\mu\nu}\, ,\quad \delta h^a{}_\mu = -\varepsilon h^a{}_\mu\, ,\quad \delta \Gamma^a{}_{b\mu} = \delta^a_b\partial_\mu\varepsilon\, ,\\
&\delta R^a{}_{b\mu\nu} = 0\, ,\quad \delta T^a{}_{\mu\nu} = -\varepsilon T^a{}_{\mu\nu}\,, \quad \delta  Q_\mu = 8\partial_\mu\varepsilon\,.
\end{split}
\end{equation*}
Here $g^M_{ab}$ is the Minkowski metric of a tangent space, $T^a{}_{\mu\nu}$ is a torsion tensor, $Q_\mu = Q^\nu{}_{\nu\mu}$ is the Weyl's vector.

The Lagrange density of the theory in a Weyl--Cartan space-time is as follows:
\begin{equation*}
\mathcal{L} = \mathcal{L}_G + \mathcal{L}_m + \frac12\sqrt{-g}\Lambda^\mu{}_{ab}\left(Q^{ab}{}_\mu -\frac14 g^{ab}Q_\mu\right)\, ,\quad \Lambda^\mu{}_{ab}g^{ab} = 0\, ,
\end{equation*}
where $\Lambda^\mu{}_{ab}$ are Lagrange multipliers, $\mathcal{L}_m$ is a Lagrange density of matter. The  Lagrange density of a gravitational field we present as follows:
\begin{equation} 
\mathcal{L}_G= \sqrt{-g}(f_0\beta^2 R + L_{R^2}+\beta^2 L_{T^2} +\beta^2 L_{Q^2} + \beta^2 L_{TQ} + L_\beta)\,, \quad f_0=\frac{1}{2\varkappa}\,,\, \varkappa = 8\pi G\,,\, c = 1.
\label{eq:LGG}
\end{equation}
It includes Lagrangians quadratic in curvature, $L_{R^2}$, and torsion, $L_{T^2}$, a Lagrangian $L_{QT}$ containing a curvature--torsion interaction, and a proper Lagrangian of the scalar field
\begin{equation*}
L_{\beta}  = l_1g^{\mu\nu}\partial_\mu\beta\partial_\nu\beta + l_2\beta\partial_\mu\beta g^{\mu\sigma} T_\sigma + l_3\beta\partial_\mu\beta g^{\mu\sigma}Q_\sigma + l_4\beta \partial_\mu\beta Q^{\mu\sigma}{}_{\sigma} + \Lambda_0 \beta^4\,.
\end{equation*}

Variational equations of the field in the Weyl-Cartan space-time have been derived by variation of the full Lagrangian density of the theory. The independent variables are tetrads, a nonholonomic connection, the scalar field and the Lagrange multipliers.

At the first stage we neglect the contributions from terms quadratic in the curvature tensor, arising from the variation of the Lagrangian $L_{R^2}$.

Due to the cumbersome, the $h-$equation and the $\Gamma-$equation are not present here, while the $\beta-$equation has the form:
\begin{equation*}
\begin{split}
\frac{\delta\mathcal{L}}{\delta\beta} =& - 2l_1\stackrel{*}{\nabla}_\mu\left(\sqrt{-g}g^{\mu\nu}\partial_\nu\beta\right) - \beta\stackrel{*}{\nabla}_\mu\left(\sqrt{-g}(l_2T^\mu + l_3Q^\mu + l_4Q^{\mu\lambda}{}_\lambda)\right)+\\
&+2\beta\sqrt{-g}\biggl(f_0R + L_{T^2} + L_{Q^2} + L_{TQ} 2\Lambda_0\beta^2\biggr) +
\frac{\delta \mathcal{L}_m}{\delta\beta} = 0\, ,
\end{split}
\end{equation*}
where $\stackrel{*}{\nabla}_\mu = \nabla_\mu + T_\mu$, $T_\mu = T^\nu{}_{\mu\nu}$. Variation of the Lagrange multipliers gives the Weyl condition on the nonmetricity tensor, $Q^{ab}{}_\mu = (1/4)g^{ab}Q_\mu$.

Variational equations are the equations of the gravitational field of the conformal theory of gravity in the Weyl--Cartan space-time with a Dirac scalar field in the tetrad formalism. These equations are now  investigated with a view to obtaining and solving an equation for the scalar Dirac field at the early stage of evolution of the universe in the absence of matter, $\mathcal{L}_m\equiv 0$.

In homogeneous and isotropic space-time the condition,  $T^a{}_{\mu\nu} = -(2/3)h^a{}_{[\mu}T_{\nu]}$,  takes place \cite{Tsamp}. By means of this condition and in view of the Weyl condition, we can express from the $\Gamma$-equation a torsion trace and Weyl vector through the Dirac scalar field $\beta$:
\begin{equation}
T_\mu = \chi_T\partial_\mu\ln\beta\,, \qquad
Q_\mu = \chi_Q\partial_\mu\ln\beta\,.
\label{eq:TQb}
\end{equation}
The coefficients in (\ref{eq:TQb}) are expressed through couple constants of the original Lagrangian.

Considering the consequence of the $h-$equation together with the $\beta-$equation, and taking into account (\ref{eq:TQb}) and the condition of homogeneity and isotropy of space, we shall get, 
\begin{equation}
\beta \ddot \beta -(k+1) (\dot \beta )^2 = 0\,, \qquad k=\frac{B}{A}\,,
\label{eq:beta}
\end{equation}
where the constants $A$ and $B$ are expressed through the parameters of the original Lagrangian.

By replacing $\beta = u^{-1/k}$ (if $k\neq 0$) or $\beta = \exp(u)$ (if $k=0$), this equation is reduced to the equation $\ddot u = 0$. Then the exact solution of equation (\ref{eq:beta}) will be:
\begin{equation}
\beta = \frac{\beta_0}{(C_1\hat t + 1)^{(1/k)}} \,,\quad (k\neq 0), \qquad  \beta = \beta_0 \exp (-C_2 \hat t)\,,\quad (k=0)\,, 
\label{eq:b(t)}
\end{equation}
where $\beta\,, C_1$ and $C_2$ are arbitrary constants of integration. The constant $\beta_0$ is the value of the scalar field in the initial time, namely, in the Planck time. For reasons of quantum field theory, this quantity should be very huge. The constants $C_1$ or $C_2$ determine the initial value of the rate of the Dirac scalar field change.

Parameter $k$ in rather complicated manner is determined by 16 of the coupling constants of the original Lagrangian (\ref{eq:LGG}). Therefore, we can always choose these constants so that one of the following conditions are fulfilled,
\begin{equation}
\frac{1}{k} = \frac{A}{B} >> 1\,, \quad \mbox{or} \quad k = 0\quad (B = 0).  
\label{eq:>>1}
\end{equation}
Thus (at least for the open universe) if one of the conditions (\ref{eq:>>1}) is valid, one can provide the necessary rapid decrease in time the scalar field value. 

The scaling-invariant theory of gravitation with scalar field in Riemann space-time has been developed in \cite{Pervush1}--\cite{Pervush3} (see also the references therein) in order to derive an alternative scenario of the evolution of the universe. In this scenario some of the observation data of modern cosmology can be explained without introducing $\Lambda$-term and without adopting the inflation hypothesis. In contradiction with this, we do not reject inflation and $\Lambda$-term. In \cite{BKF:Kazan},  \cite{BKF:MathSb}, \cite{FrKouch} we have expressed the hypothesis that the dark energy is determined by the value of the Dirac scalar field and determined by the term $\Lambda_0 \beta^4$ of the Lagrangian. 

Thus our result can explain the rapid decrease in time of the dark energy (the energy of physical vacuum). The decrease of dark energy according to (\ref{eq:b(t)}) is much more intense than the corresponding decrease, which can be carried out in Poincare gauge theory of gravity \cite{MinkGarKud}, \cite{Mink2}, in which the effective cosmological term is defined by the torsion tensor.

We point out that the ultra-rapid decrease of the energy of physical vacuum according to the law (\ref{eq:b(t)}) occurs only prior to the Friedmann era evolution of the universe. According the scenario of inflation, the birth of the rest masses of elementary particles begins in the last period of inflation, and further evolution of the universe is determined not by a scalar field, but mainly by the born ultrarelativistic matter and the radiation interacting with it. Solution (\ref{eq:>>1}) requires a suitable modification in this case.

\end{document}